\documentclass[aps,pra,showpacs,groupedaddress,twocolumn]{revtex4}
\usepackage{amsfonts}
\usepackage{graphicx}
\usepackage{latexsym}
\usepackage{makeidx}
\usepackage{amsmath}
\usepackage{amssymb}

\input{tcilatex}

\begin{document}
\title{AMENDART in Markovian circuit QED}
\author{A. V. Dodonov}
\affiliation{Instituto de F\'{\i}sica, Universidade de Bras\'{\i}lia, Caixa Postal 04455,
70910-900 Bras\'{\i}lia, DF, Brazil}
\pacs{42.50.Pq, 42.50.Ct, 03.65.Yz, 42.50.Hz}

\begin{abstract}
We study the cavity field's and atomic \emph{asymptotic mean excitation
numbers due to anti-rotating term} (AMENDART) in the circuit Quantum
Electrodynamics (circuit QED) system, composed of a two-level atom and a
single cavity field mode, subject to Markovian damping and dephasing
mechanisms. We show that the AMENDART are above the thermal values, and
their behavior is analyzed analytically and numerically for typical
parameters in circuit QED implementations described by the Rabi Hamiltonian.
We point out that \textquotedblleft parasitic elements\textquotedblright\,
such as other cavity modes or eventual off-resonant atoms, also contribute
substantially to AMENDART.
\end{abstract}

\maketitle

\section{Introduction}

\label{1}

Cavity Quantum Electrodynamics (cavity QED) is one of the most active areas
of research in quantum optics and quantum information nowadays. It deals
with the light-matter interaction for electromagnetic (EM) fields with a few
photons and a small number of real or artificial atoms confined in optical
and microwave resonators. It can be implemented in a variety of
architectures, such as atomic ensembles, Rydberg atoms, quantum dots,
superconducting circuits, Bose-Einstein condensates, polar molecules,
trapped ions, etc \cite%
{Walmsley,Walmsley1,Walmsley2,Walmsley3,Walmsley4,Harro}. Here we focus on
the cavity QED realizations in superconducting circuits, the area known as
\emph{circuit QED} \cite{Blais,refe}, where the one-photon--one-atom
interaction is currently realized with different types of artificial atoms,
and interaction between different atoms, placed deterministically inside the
cavity, is realized by means of the cavity field acting as the quantum bus.
Some advantages of using superconducting circuits are the possibility of
engineering the properties of the artificial atoms, such as the energy
levels configuration and the atom-field coupling, rapid tuning of the
cavity's and atom's frequencies and the freedom in fabricating several
artificial atoms at specific locations within the cavity. Besides studying
fundamental problems in quantum mechanics, such as entanglement,
dissipation, decoherence and the measurement problem, circuit QED is also an
important tool for testing basic quantum algorithms \cite{dicarlo},
engineering nonclassical states of light and matter \cite%
{Walmsley,Walmsley4,Har2}, achieving the strong and ultra-strong
light-matter coupling \cite{Hows,Hows1}, implementing rapid time-dependent
phenomena \cite{Libe,Libe1,Liberato,Lukin,Lukin1,v7,v9,duty} and novel
technologies for quantum information processing \cite{Walmsley4,QIP}, etc.

The simplest Hamiltonian, deduced from first principles, that describes the
interaction between a two-level atom and a single mode of the quantized
electromagnetic field in a cavity is the \emph{Rabi Hamiltonian} (RH) \cite%
{Rabi}. It reads $(\hbar =1)$%
\begin{equation}
H_{R}=n+\Omega E+g\,p\sigma _{y},  \label{HR}
\end{equation}%
where we set the cavity frequency to $\nu =1$, the atomic transition
frequency is $\Omega $ and $g$ is the atom-field coupling constant. The
cavity field quadrature operators are $x=(a^{\dagger }+a)/\sqrt{2}\ $and $%
p=i(a^{\dagger }-a)/\sqrt{2}$, $a$ and $a^{\dagger }$ are the bosonic
annihilation and creation operators, satisfying the commutation relation $%
[a,a^{\dagger }]=1$, and $n=a^{\dagger }a$ is the photon number operator.
The atomic operators are $E=\left\vert e\right\rangle \left\langle
e\right\vert $, $\sigma _{x}=\sigma _{-}+\sigma _{+}$, $\sigma _{y}=i(\sigma
_{-}-\sigma _{+})$ and$~\sigma _{z}=E-|g\rangle \langle g|$,\thinspace where
$\sigma _{+}=\left\vert e\right\rangle \left\langle g\right\vert $ and $%
\sigma _{-}=\sigma _{+}^{\dagger }$, $\left\vert g\right\rangle $ and $%
\left\vert e\right\rangle $ denoting the ground and excited atomic states,
respectively. The exact diagonalization of RH has not been achieved yet, so
to obtain analytical results one usually performs the Rotating Wave
Approximation (RWA) by neglecting the \textit{anti-rotating term }(ART) $%
(a\sigma _{-}+a^{\dagger }\sigma _{+})$, responsible for simultaneous
creation/annihilation of one photon and one atomic excitation \cite%
{Scully,Schleich}. In this case one gets the celebrated \emph{%
Jaynes-Cummings Hamiltonian} (JCH) \cite{JC,JC1}, which has a simple
analytical solution \cite{Scully,Schleich,Harro} that led to prediction of
several pure quantum effects, many of them already verified experimentally
\cite{Harro,Walmsley,Walmsley1,Walmsley2,Har1,Har2,Har3,refe}.

The RH or JCH alone do not describe the current situation in circuit QED
because of dissipation and decoherence arising from the system interaction
with cavity's and atomic environments. So instead of the Schr\"{o}dinger
equation one has to use the quantum master equation for the density operator
$\rho $, whose general form is \cite{Scully,Schleich,Breuer,Carm}%
\begin{equation}
\frac{d}{dt}{\rho }+i[H,\rho ]=\mathcal{L}\rho ,  \label{ssme1}
\end{equation}%
where $H$ is some effective Hamiltonian and the superoperator $\mathcal{L}$
is the Liouvillian that accounts for the influence of the environment. In
most situations the Liouvillian consists of three parts, $\mathcal{L}=%
\mathcal{L}_{\kappa }+\mathcal{L}_{\lambda }+\mathcal{L}_{\gamma }$, where
the superoperator $\mathcal{L}_{\kappa }$ ($\mathcal{L}_{\lambda }$)
describes the cavity (atom) damping by the thermal reservoir with the mean
photon number $\bar{n}$, and $\kappa $ ($\lambda $) is the cavity (atom)
relaxation rate that can be determined experimentally \cite%
{Walmsley,Harro,refe,refe1,Har1,Har2,Har3}. Another common source of
decoherence in superconducting circuits is the pure atomic dephasing
represented by $\mathcal{L}_{\gamma }$, that arises due to low-frequency $%
1/f $ noise \cite{Clarke}, and $\gamma $ denotes the pure dephasing rate
\cite{Carm,Makhlin}. The preferred theoretical pick to describe the majority
of experiments, owing to its simplicity and wide applicability range, is the
\textquotedblleft standard master equation\textquotedblright\ (SME) \cite%
{Harro,Har1,Blais,refe,refe1}
\begin{eqnarray}
\mathcal{L}_{\kappa } &=&\kappa (\bar{n}+1)\mathcal{D}[a]+\kappa \bar{n}%
\mathcal{D}[a^{\dagger }]  \label{kik1} \\
\mathcal{L}_{\lambda } &=&\lambda (\bar{n}+1)\mathcal{D}[\sigma
_{-}]+\lambda \bar{n}\mathcal{D}[\sigma _{+}] \\
\,\mathcal{L}_{\gamma } &=&\frac{\gamma }{2}\mathcal{D}[\sigma _{z}],
\label{kik2}
\end{eqnarray}%
that can be deduced microscopically by making the Born-Markovian
approximation on the system-reservoir interaction \cite{Carm}, where the so
called Lindblad superoperator%
\begin{equation}
\mathcal{D}[\Phi ]\rho \equiv \frac{1}{2}\left( 2\Phi \rho \Phi ^{\dagger
}-\Phi ^{\dagger }\Phi \rho -\rho \Phi ^{\dagger }\Phi \right)
\end{equation}%
preserves the hermiticity, normalization and positivity of $\rho $ \cite%
{Breuer}.

Within the quantum trajectories approach a Markovian reservoir represented
by a Lindblad kernel can be viewed as a \textquotedblleft
detector\textquotedblright\ performing continuous measurements on the system
\cite{Breuer,prel,engi}. Considering, for instance, $\mathcal{L=}\chi
\mathcal{D}[\Phi ]$ and rewriting (\ref{ssme1}) as $d\rho /dt-L\rho =R\rho $%
, where $R\rho \equiv \chi \Phi \rho \Phi ^{\dagger }$ and $L\rho \equiv
-i[H,\rho ]-\{\Phi ^{\dagger }\Phi ,\rho \}\chi /2$, one may check that the
formal solution for $\rho (t)$ is \cite{Carm}
\begin{eqnarray}
\rho (t) &=&\sum_{m=0}^{\infty
}\int_{0}^{t}dt_{m}\int_{0}^{t_{m}}dt_{m-1}\cdots \int_{0}^{t_{2}}dt_{1}%
\tilde{\rho}_{m} \\
\tilde{\rho}_{m} &\equiv &e^{L(t-t_{m})}Re^{L(t_{m}-t_{m-1})}\cdots
Re^{Lt_{1}}\rho \left( 0\right) .
\end{eqnarray}%
This procedure decomposes the quantum dynamics contained in the master
equation into an infinity of \emph{quantum trajectories}, where each
occurrence of $R$ corresponds to an instantaneous quantum jump of the
system's state \cite{engi}, and the exponentials $\exp [L_{0}(t_{k}-t_{k-1})]
$ describe the non-unitary system evolution between the quantum jumps. Thus $%
\tilde{\rho}_{m}$ can be interpreted as the evolved (non-normalized) density
operator conditioned to quantum jumps at times $t_{1},t_{2}\ldots t_{m}$,
and its trace gives the probability of this particular sequence. Such an
interpretation describes the \emph{Markovian} dissipation as continuous
measurement of the system by the reservoir\ \cite{prel}, where the quantum
jumps describe the \textquotedblleft clicks\textquotedblright\ of the
\textquotedblleft detector\textquotedblright\ \cite{engi}, and the
exponential superoperators describe the system evolution without
\textquotedblleft clicks\textquotedblright\ but still under continuous
monitoring \cite{Breuer,Carm}.

The JCH shows good agreement with experiment and with numerical calculations
in the majority of problems of practical interest \cite{Har3}. However, in
some occasions the ART becomes important, so many studies investigated the
range of validity of RWA and new physics beyond RWA (see references in \cite%
{CAMOP}). In particular, in \cite{Werlang1} was shown that the combined
action of the anti-rotating term and Markovian dissipation leads to
incoherent photon creation from vacuum, so the asymptotic mean photon number
is slightly above the thermal value. A detailed analysis of this phenomenon
was performed in \cite{CAMOP}, in \cite{Werlang3} the cavity emission rate
due to ART was calculated, and the relation of ART to the entropy operator
was described in \cite{Kurcz2}. Moreover, in the study \cite{2atoms} the
errors in the zero-excitation state preparation due to ARTs in two-atom
Markovian cavity QED system were estimated, while the creation of transient
entanglement between two atoms due to ARTs was reported in \cite{Jing}.

Here we study analytically and numerically the cavity field's and atomic
\emph{asymptotic mean excitation numbers due to anti-rotating term}
(AMENDART) for the current parameters in circuit QED, taking into account
common \textquotedblleft parasitic\textquotedblright\ elements, such as
other cavity modes and off-resonant two-level atoms, showing that in many
cases these elements contribute significantly to the generation of atomic
and cavity field's excitations due to ART.

\section{Analytical and numerical results}

\label{2}

In this section we analyze analytically and numerically the asymptotic mean
numbers of photons and atomic excitations, $\left\langle n\right\rangle $
and $\left\langle E\right\rangle $, respectively, in the circuit QED setup
described by the master equation (\ref{ssme1}) with $H=H_{R}$ and
dissipative kernels (\ref{kik1})-(\ref{kik2}). Since RH cannot be integrated
exactly, to obtain approximate analytical results we perform the $k$-photons
approximation (denoted below by the subindex $k$) by assuming that at most $%
k $ photons are present in the cavity. The asymptotic values are evaluated
by equaling to zero the left-hand-side of the resulting Heisenberg equations
of motion. Under the $1$-photon approximation, valid for $\left\langle
n_{1}\right\rangle \ll 1$, for zero-temperature damping reservoirs ($\bar{n}%
=0$) we get simple formulae for the AMENDART \cite{CAMOP}%
\begin{eqnarray}
\left\langle n_{1}\right\rangle &=&\frac{G}{T}\left( 2G+\lambda s\right)
\label{r1} \\
\langle E_{1}\rangle &=&\frac{G}{T}\left( 2G+\kappa s\right) ,
\end{eqnarray}%
where $G=g^{2}\Gamma $, $\Gamma =\gamma +(\kappa +\lambda )/2$, $s=\Delta
^{2}+\Gamma ^{2}$, $T=2G[\alpha (\kappa +\lambda )+2G]+\lambda \kappa \beta $%
, $\alpha =s+2\Omega $ and $\beta =\alpha ^{2}-4\Omega ^{2}$. Here $\Delta
\equiv \Omega -1$ is the atom-cavity detuning and $\left\langle \cdots
\right\rangle $ denotes the asymptotic value. One may also obtain another
useful expression for estimating the correlation between the atom and the
field%
\begin{equation}
\left\langle (n\sigma _{z})_{1}\right\rangle =\frac{\lambda }{\lambda
+\kappa }\left( \left\langle E_{1}\right\rangle -\left\langle
n_{1}\right\rangle \right) .  \label{r3}
\end{equation}%
For higher orders approximations the formulae are not so compact, so below
we shall present only the numerical results for the $2$-photons
approximation.

In many practical situations the circuit QED system contains other elements,
apart from the selected cavity mode and the artificial atom. In particular,
the atom always interacts with other cavity modes, usually neglected due to
their strong detuning from the atomic transition frequency and because
initially they are not populated by photons. In ordinary cavities realized
as coplanar waveguide resonators the modes are regularly spaced in
frequency, and if the atom is fabricated at one end of the resonator, then
all the cavity modes have a voltage antinode at the atom's position \cite%
{Blais}. To check whether other cavity modes are relevant for the phenomenon
of photon generation due to the ART, we suppose that, besides the selected
cavity mode (the \textquotedblleft true mode\textquotedblright ), the atom
also interacts with another cavity mode with frequency $\tilde{\nu}$ and the
corresponding coupling constant $\sqrt{\tilde{\nu}}g$ -- a \textquotedblleft
parasitic mode\textquotedblright , whose variables are denoted with tilde.
In this case the total Hamiltonian becomes \cite{Scully,Schleich}%
\begin{equation}
H=H_{C}\equiv H_{R}+\tilde{\nu}\tilde{n}+\sqrt{\tilde{\nu}}g\,\tilde{p}%
\sigma _{y},
\end{equation}%
and assuming that the cavity quality factor is the same for the two modes,
the corresponding modification in the Liouvillian is%
\begin{equation}
\mathcal{L}_{C}=\mathcal{L}+\tilde{\nu}\kappa \mathcal{D}[\tilde{a}].
\end{equation}%
Based on common experimental conditions, we assume that initially the
parasitic mode is in the vacuum state, and consider two plausible scenarios.
a) The true mode ($\nu =1$) is the fundamental mode of the half-wave
resonator, and the parasitic mode corresponds to the first harmonic full
wavelength resonance, $\tilde{\nu}=2$. b) The opposite scenario: the true
mode ($\nu =1$) corresponds to the first harmonic full wavelength resonance,
and the parasitic mode is the half-wave fundamental mode, $\tilde{\nu}=1/2$.
Below we distinguish between these two cases by indicating the value of $%
\tilde{\nu}$.

Furthermore, in some circuit QED setups two almost identical artificial
2-level atoms are fabricated within the same cavity to perform two-atoms
quantum gates, where the cavity mode plays the role of a quantum bus. When
one is interested in performing only one-atom quantum gates, the second atom
is effectively \textquotedblleft turned off\textquotedblright\ by tuning its
transition frequency $\tilde{\Omega}$ far apart from the cavity mode's and
the other atom's frequencies. So we study the scenario where the
\textquotedblleft connected\textquotedblright\ atom\ is the
\textquotedblleft true atom\textquotedblright\ and the \textquotedblleft
disconnected\textquotedblright\ one acts as the \textquotedblleft parasitic
atom\textquotedblright\ (its variables are denoted with tilde), so the total
Hamiltonian and the Liouvillian become \cite{2atoms}%
\begin{equation}
H=H_{A}\equiv H_{R}+\tilde{\Omega}\tilde{E}+g\,p\tilde{\sigma}_{y}
\end{equation}%
\begin{equation}
\mathcal{L}_{A}=\mathcal{L}+\lambda \mathcal{D}[\tilde{\sigma}_{-}]+\frac{%
\gamma }{2}\mathcal{D}[\tilde{\sigma}_{z}],
\end{equation}%
where we assumed that the coupling constants and the dissipative rates are
the same for the both atoms. We shall consider two scenarios: c) the
parasitic atom transition frequency is significantly smaller than the cavity
mode frequency, $\tilde{\Omega}=0.2$, or d) is significantly larger, $\tilde{%
\Omega}=1.8$.

\begin{figure}[th]
\includegraphics[width=.5\textwidth]{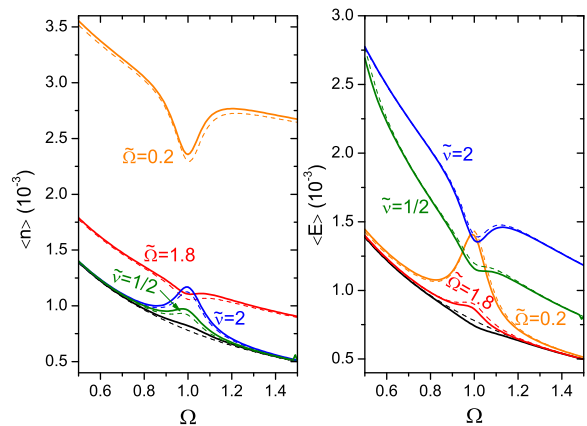}\hspace{2pc}%
\caption{\label{w}AMENDART as function of atomic transition frequency
for parameters $\protect\kappa =\protect\lambda =10^{-6}${} and $\protect%
\gamma =\protect\lambda /4$ in the presence of parasitic elements, whose
parameters are indicated in the plots. Dashed lines correspond to the $1$%
-photon approximation, and the solid ones -- to
the $2$-photons approximation.}
\end{figure}

\begin{figure}[ht]
\includegraphics[width=.5\textwidth]{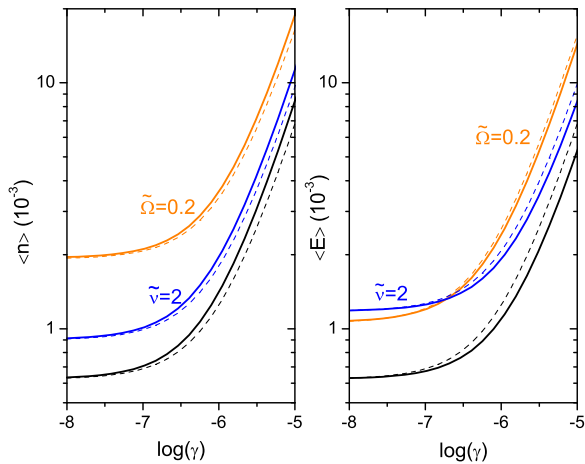}\hspace{2pc}%
\caption{\label{q}AMENDART as function of $\protect\gamma $ for
parameters $\protect\kappa =\protect\lambda =10^{-6}${}, $\Omega =1$ in the
presence of parasitic elements, whose parameters are indicated in the plots.
The solid lines correspond to the $2$-photons approximation,
and the dashed ones -- to the $1$-photon approximation.}
\end{figure}

\begin{figure}[ht]
\includegraphics[width=.5\textwidth]{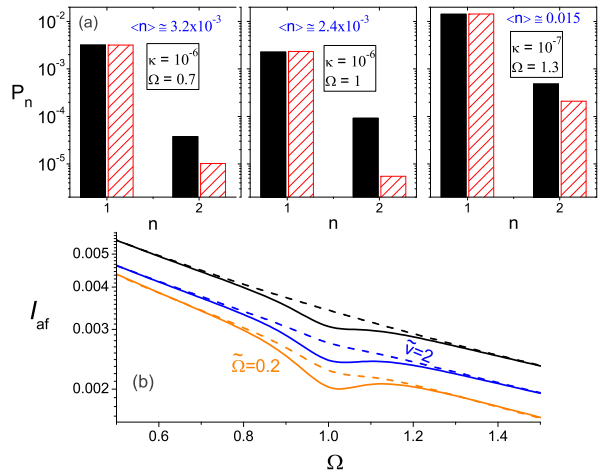}\hspace{2pc}%
\caption{\label{distr}a) Photon number distribution for parameters $\tilde{%
\Omega}=0.2$, $\protect\lambda =10^{-6}$, $\protect\gamma =\protect\lambda %
/4 $ and $\protect\kappa $ and $\Omega $ as indicated in the plots. Filled
black bars describe the photon distribution due to AMENDART, and the red
bars with sparse pattern correspond to the thermal distribution with the
same $\left\langle n\right\rangle $. b) Quantum mutual information $I_{af}$ for the asymptotic state with and without the
parasitic elements.}
\end{figure}

\begin{figure}[ht]
\includegraphics[width=.5\textwidth]{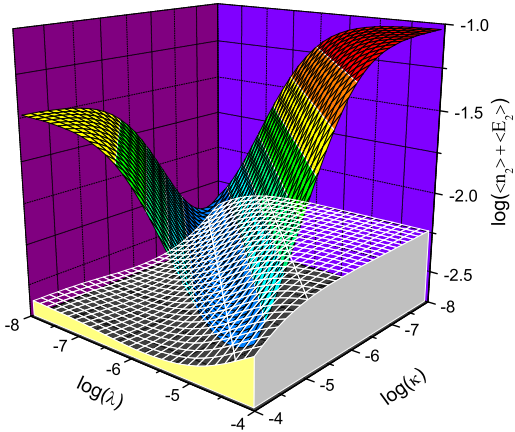}\hspace{2pc}%
\caption{\label{kl}Logarithm of the total mean excitation number $\log
(\left\langle n_{2}\right\rangle +\left\langle E_{2}\right\rangle )$ as
function of $\log (\protect\kappa )$ and $\log (\protect\lambda )$ in the
presence of the parasitic atom, for parameters $\protect\gamma =\protect%
\lambda /4$ and $\tilde{\Omega}=0.2$. The colored curve corresponds to
out-of-resonance case, $\Omega =0.7$, and the white one to the resonance, $%
\Omega =1.$}
\end{figure}

Below we study numerically the AMENDART of the true cavity mode, $%
\left\langle n\right\rangle $, and the true atom, $\left\langle
E\right\rangle $, as function of the system parameters under the $1$- and $2$%
- photon approximations. We verified that in the \emph{weak coupling limit},
$g\ll 1$, these quantities grow up approximately quadratically in $g$, so in
the figures \ref{w} -- \ref{kl} we set $g=5\times 10^{-2}$, a value that can
be achieved in some circuit QED systems and is within the weak coupling
limit. Moreover, as we are interested in the lower bounds for the mean
excitation numbers, we set the reservoir temperatures to $0\,\mathrm{K}$, $%
\bar{n}=0$. This simplification does not restrict the scope of our analysis,
since we verified that for small but finite temperatures ($\bar{n}\ll 1$)
the cavity's and atomic excitations due to finite temperature are
approximately added to the AMENDART (data not shown).

In the figure \ref{w} we show how $\left\langle n\right\rangle $ and $%
\left\langle E\right\rangle $ depend on the true atom transition frequency $%
\Omega $ in the presence of parasitic elements for parameters indicated in
the caption, that can be achieved in the near future experiments. The dashed
lines correspond to the $1$-photon approximation and the solid lines -- to
the $2$-photons approximation. We see that for the chosen parameters the $1$%
-photon approximation is quite accurate, and out of resonance the AMENDART
decrease as function of $\Omega $. In the figure \ref{q} we perform a
similar analysis as function of the pure dephasing rate, from which we see
that the AMENDART grow up as function of $\gamma $, and the $1$-photon
approximation loses its accuracy as $\left\langle n\right\rangle $ (or $%
\gamma $) increases. Remarkably, in these cases the parasitic elements \emph{%
cannot} be neglected, since they contribute substantially to the AMENDART.
Therefore, if one wants to know precisely the values of AMENDART in circuit
QED systems, one has to take into account the parasitic elements, even when
they are far detuned and initiated in their respective ground states.

Despite a finite asymptotic photon population in the true cavity mode, it
cannot be associated to an effective temperature reservoir \cite{Klimov},
since the resulting distribution is significantly different from the thermal
one. This is illustrated in the figure \ref{distr}a, where we show the
probability $P_{n}$ of having $n$ photons in the cavity, obtained under the $%
2$-photons approximation in the presence of the parasitic atom for
parameters indicated in the caption (filled black bars). For comparison we
also show the thermal distribution with the same mean photon number (red
bars with sparse pattern), whereby one can see that the two distributions
are quite different. Besides, there is a correlation between the
\textquotedblleft true field\textquotedblright\ and the \textquotedblleft
true atom\textquotedblright\ in the asymptotic state, as quantified by the
\textit{quantum mutual information}, that measures the \textit{total} amount
of correlation in a bipartite quantum state \cite{Nielsen}%
\begin{equation}
I_{af}=S\left( \rho _{a}\right) +S\left( \rho _{f}\right) -S\left( \rho
_{af}\right) .
\end{equation}%
Here $S\left( \rho _{k}\right) =-\mathrm{Tr}[\rho _{k}\log _{2}(\rho _{k})]$
is the von Neumann entropy of the $k$th subsystem whose dynamics is
described by the reduced asymptotic density matrix $\rho _{k}$ with $%
k=\{atom,field,atom-field\}$. $I_{af}$ is shown in the figure \ref{distr}b
as function of $\Omega $ in the absence of the parasitic elements (black
lines), and when the \textquotedblleft parasitic atom\textquotedblright\
(orange lines) or \textquotedblleft parasitic mode\textquotedblright\ (blue
lines) are present for parameters $\kappa =\lambda =10^{-6}$ and $\gamma
=\lambda /4$. The dashed (solid) lines correspond to the $1$ ($2$)-photon
approximation, and the results for the $1$-photon approximation in the
absence of the parasitic elements can be obtained analytically with the aid
of equations (\ref{r1})-(\ref{r3}). One can see that asymptotically the
atom-field system is correlated, $I_{af}\neq 0$, and the mutual information
decreases in the presence of the parasitic elements, since in this case they
acquire some information about the system of interest.

One can also notice from the right plot of figure \ref{distr}a that when the
true atom is out of resonance, $\left\langle n\right\rangle $ is larger for
smaller $\kappa $. This is better depicted in the figure \ref{kl}, where we
show the total asymptotic mean excitation number under the $2$-photons
approximation, $\left\langle n_{2}\right\rangle +\left\langle
E_{2}\right\rangle $, as function of damping rates under the resonance
(sparse white curve) and out-of-resonance (filled colored curve) conditions
in the presence of the parasitic atom, with the system parameters indicated
in the caption. One can see that at the resonance $\left\langle
n_{2}\right\rangle +\left\langle E_{2}\right\rangle $ does not depend
strongly on the damping rates, contrary to out-of-resonance case, where it
is smaller when $\kappa $ and $\lambda $ are approximately equal.


\section{Discussion and generalizations}

\label{3}

The phenomenon of excitations generation above the thermal values occurs due
to the combined action of the ART in the RH and the Markovian approximation
for the reservoirs \cite{CAMOP}. The ART describes continuous spontaneous
creation from vacuum and annihilation of virtual atomic and photonic
excitations \cite{Scully,Schleich}. Under the Markovian approximation, the
reservoir correlation time is very short compared to the time scale for a
significant change in the system \cite{Carm}, i.e. the time needed to
annihilate excitations. Therefore, once the excitations are spontaneously
created, a fraction of them decays to the reservoir due to atomic and/or
cavity damping before they can be annihilated, so the system coherence is
destroyed and the initial zero-excitation state cannot be restored. From the
point of view of quantum trajectories \cite{Breuer,Carm,prel}, the
continuous monitoring of the system by the Markovian environment promotes
the virtual excitation into real ones due to the continuous change of
system's state, while a fraction of the excitations is destroyed due to
destructive measurements, thereby asymptotically nonzero photon and atomic
populations are built in the system. The pure atomic dephasing increases the
number of created excitations \cite{Werlang1} because the dephasing
reservoir performs nondestructive measurement of the atomic state \cite{Carm}%
, so the virtual excitations are promoted into real ones nondestructively.
Besides, the Markovian pure atomic dephasing can be attributed to random
changes of the atomic transition frequency \cite{refe,Makhlin,Werlang1},
which promote the virtual excitations into real ones due to the modification
of the system ground state, analogously to the dynamical Casimir effect \cite%
{Liberato,Book,Lukin,Lukin1}. It is worth noting that, by definition, the
pure dephasing reservoir always has a finite temperature \cite{Carm},
whereby it can perform nondestructive measurements of the atomic state, so
the substantial amplification of system excitations due to pure Markovian
dephasing is not so puzzling.

The phenomenon of nonzero asymptotic photon population also occurs for a
single dissipative cavity if one takes into account the anti-rotating
cavity-reservoir interaction. Suppose we isolate a single constituent of the
reservoir (e.g. 2-level atom, harmonic oscillator, etc) and treat it as an
ancilla; then, one performs the standard Born-Markovian master equation
treatment over the remaining reservoir constituents. Thereby we obtain a
Markovian master equation for the cavity-ancilla system, including the
anti-rotating interaction. By performing the asymptotic analysis of the
previous section and tracing over the ancilla variables at the end, one
would end up with a nonzero photon population due to the virtual
cavity-ancilla excitations being promoted into real ones by the Markovian
reservoir. Another way of arriving at this conclusion is to use the
simplified phenomenological approach. The most general master equation for a
single cavity field mode \cite{Dekker,Me}, preserving the normalization and
hermiticity of the statistical operator $\rho $ and containing \emph{only
bilinear}\/ forms of operators $x$ and $p$ is given by the equation (\ref%
{ssme1}) with the effective Hamiltonian%
\begin{equation}
H=H_{0}+\frac{\mu }{4}\left\{ x,p\right\} ,  \label{Hmu}
\end{equation}%
where $H_{0}=n$ is the cavity Hamiltonian (recalling that the cavity
frequency is set to $\nu =1$), and the damping superoperator is $\mathcal{L}=%
\mathcal{L}_{\kappa }^{\prime }$%
\begin{eqnarray}
\mathcal{L}_{\kappa }^{\prime }\rho &=&\frac{i\kappa }{4}\left( \left[
p,\left\{ x,\rho \right\} \right] -\left[ x,\left\{ p,\rho \right\} \right]
\right) -D_{p}\left[ x,\left[ x,\rho \right] \right]  \nonumber \\
& -& D_{x}\left[ p,\left[
p,\rho \right] \right] +D_{z}\left( \left[ x,\left[ p,\rho \right] \right] +%
\left[ p,\left[ x,\rho \right] \right] \right) ,  \label{MEQ}
\end{eqnarray}%
with $\mu $, $\kappa $, $D_{x}$, $D_{p}$ and $D_{z}$ being arbitrary
time-independent coefficients under the Markovian approximation. The
condition
\begin{equation}
D_{p}D_{x}-D_{z}^{2}\geq (\kappa /4)^{2}  \label{restrDxpz}
\end{equation}%
guarantees that the positivity of the statistical operator is preserved for
all times and for \emph{any physically admissible} initial state. This is
the necessary and sufficient condition (together with conditions $D_{x}\geq
0 $ and $D_{p}\geq 0$) of reducibility of the superoperator (\ref{MEQ}) to
the Lindblad form \cite{DOM85}. The standard master equation (\ref{kik1})
corresponds to the choice $D_{p}=D_{x}=\kappa \left( 1+2\bar{n}\right) /4$
and $D_{z}=\mu =0$, which satisfies the inequality (\ref{restrDxpz}).

One can verify \cite{CAMOP} that asymptotically the vacuum state (with mean
values $\langle x^{2}\rangle =\langle p^{2}\rangle =1/2$, $\langle \left\{
x,p\right\} \rangle =0$) is achieved for the coefficients
\begin{equation}
D_{p}^{\prime }=(\kappa +\mu )/4,\quad D_{x}^{\prime }=(\kappa -\mu
)/4,\quad D_{z}^{\prime }=0.
\end{equation}%
Using the inequality (\ref{restrDxpz}) one obtains $\kappa ^{2}-\mu ^{2}\geq
\kappa ^{2}$, whose solution is $\mu =0$. Therefore the vacuum state can be
achieved \emph{only} for the coefficients $D_{x}^{\prime }=D_{p}^{\prime
}=\kappa /4 $ and $\mu =D_{z}^{\prime }=0$, which correspond to the standard
master equation (\ref{kik1}) at zero temperature. However, the SME is
deduced microscopically by making the RWA on the cavity-reservoir
interaction \cite{Carm,Scully,Schleich}, i.e. by neglecting the
anti-rotating terms responsible for simultaneous creation of one virtual
photon and a virtual reservoir excitation. Therefore, if the anti-rotating
cavity-reservoir interactions are taken into account, under the Markovian
approximation the asymptotic mean photon number in the dissipative cavity is
larger than zero. Finally, if one uses a single dissipative kernel (\ref{MEQ}%
) together with the Hamiltonian\ (\ref{Hmu}) to describe the circuit QED
system, with $H_{0}=H_{R}$, under the $1$-photon approximation one gets (for
$g\neq 0$)%
\begin{eqnarray}
\left\langle n_{1}\right\rangle &=&\frac{1}{2}\left( 1-\frac{1}{2}\frac{%
\kappa }{D_{xp}+2g^{2}D_{x}\phi }\right) >0 \\
\langle E_{1}\rangle &=&\frac{1}{2}\left( 1-\frac{1}{2}\frac{\kappa \Omega
\phi \nu _{+}D_{xp}/D_{p}}{D_{xp}+2g^{2}D_{x}\phi }\right) >0,
\end{eqnarray}%
where $D_{xp}=D_{x}+D_{p}$, $\nu _{\pm }=\left( 1\pm 2D_{z}\right) $ and $%
\phi =\left[ \Omega ^{2}+4D_{x}^{2}+\nu _{+}\nu _{-}D_{x}/D_{p}\right] ^{-1}$%
. This demonstrates that regardless of the exact form of the master equation
and the amount of dissipative channels, under the \emph{Markovian
approximation} the asymptotic mean photon and the atomic excitation numbers
are always greater than zero, so the vacuum state is never achieved exactly.

\section{Summary}

In summary, we studied the behavior of the cavity field's and atomic
asymptotic mean excitation numbers due to anti-rotating term (AMENDART) for
typical parameters of the Markovian circuit QED system, showing that
\textquotedblleft parasitic elements\textquotedblright\ (such as other
cavity modes and off-resonant atoms) contribute to these quantities, which
are typically of the order of $10^{-3}$ when the atom-field coupling
constant is within a few percents of the cavity resonant frequency. This
result implies that whenever one uses a Markovian master equation to
describe the circuit QED system, there is a small intrinsic uncertainty in
the mean photon number and the atomic excitation probability that originates
from the anti-rotating term in the light-matter interaction Hamiltonian.

\begin{acknowledgments}
The author acknowledges partial financial support by DPP/UnB (Bras\'{\i}%
lia, DF, Brazil), edital 04/2010.
\end{acknowledgments}

\end{document}